\documentclass[pra,twocolumn,groupedaddress]{revtex4}
\usepackage{graphics,epsfig,mathrsfs}

\begin{document}

\title{Optical signatures of quantum phase transitions in a light-matter system}

\author{Timothy C. Jarrett, Alexandra Olaya-Castro and Neil F. Johnson}

\affiliation{Centre for Quantum Computation and Department of
Physics, Clarendon Laboratory, University of Oxford, Oxford, OX1
3PU, United Kingdom}

\date{\today}

\begin{abstract}
Information about quantum phase transitions in conventional condensed matter
systems, must be sought by probing the matter system itself. By contrast, we
show that mixed matter-light systems offer a distinct advantage in that the
photon field carries clear signatures of the associated quantum critical
phenomena. Having derived an accurate, size-consistent Hamiltonian for the
photonic field in the well-known Dicke model, we predict striking behavior of
the optical squeezing and photon statistics near the phase transition. The
corresponding dynamics resemble those of a degenerate parametric
amplifier. Our findings boost the motivation for exploring exotic quantum phase
transition phenomena in atom-cavity, nanostructure-cavity, and
nanostructure-photonic-band-gap systems.

\vskip0.2in
\end{abstract}

\maketitle

There are several theoretical models which are currently attracting attention, based on the possible insights that they offer into the nature of Quantum Phase Transitions (QPTs). One of these is the Dicke model which was originally developed in quantum optics, together with its recent generalizations \cite{CMP,Brandes,Reslen}. In practice, such exotic quantum phenomena can only be studied experimentally if the system's many-body quantum state can be probed in some way \cite{Carmichael}. Unfortunately in condensed matter systems, such probing is typically indirect since one cannot `see' many-body quantum spin states. 

Here we predict that in light-matter systems approximating to the Dicke model -- such as atom-cavity and nanostructure-cavity systems \cite{CMP,Brandes,Reslen,Carmichael} -- the statistical properties of the photon field offer direct and striking signatures of the quantum critical phenomena underlying a QPT. Our results are based on an accurate, size-consistent calculation of the statistical properties of the photon field in the Dicke model, and help motivate the exploration of such exotic quantum phenomena in atom-cavity, nanostructure-cavity, and nanostructure-photonic-band-gap systems \cite{CMP,Brandes,Reslen,Carmichael}. In addition to the results themselves, our theoretical approach presents a number of distinct features over previous works \cite{CMP,Brandes,Reslen,Carmichael}: (1) we avoid using canonical perturbation schemes and projection methods, which can suffer from inconsistent size-dependencies as one approaches the thermodynamic limit \cite{Becker}. Instead we adopt a similar renormalization-like scheme to Ref. \cite{Reslen}, but take the opposite viewpoint by renormalizing the dynamics of the photon field as opposed to the matter system.  (2) Our approach shows the direct connection between the Dicke model and a degenerate parametric optical amplifier. (3) In addition to explicitly reproducing the correct scaling near the critical point, we are able to show that striking signatures arise in a number of key statistical properties associated with the photon field.
(4) We are able to show the optical manifestation of a quasi-integrable to quantum chaotic transition near the QPT. 

The Dicke model describes the interaction between a single-mode photon field and $N$ non-interacting two-level systems
\cite{CMP,Reslen}:
\begin{eqnarray}
 H &=& a^\dag a + \epsilon J_z +
      \frac{\lambda}{2\sqrt{N}}(a + a^\dag)J_x,
\end{eqnarray}
where $J_z=\frac{1}{2}\sum_{i=1}^N \sigma^z_i$ and  
$J_x=\frac{1}{2}\sum_{i=1}^N (\sigma^+_i + \sigma_{i}^{-})$ 
are the collective angular momentum operators, and the operators 
$a, a^\dag$ and $\sigma^{\pm}_j, \sigma^z_j$ correspond to
the photon field and two-level atom $i$ respectively. This
model exhibits a phase transition at both zero and finite temperature
\cite{CMP,Reslen}. Employing the cumulant expansion method \cite{Reslen,Becker}, we here choose to eliminate the degrees of freedom in
the matter subsystem and hence derive a size-consistent effective Hamiltonian for the 
{\em photon} field.  
Consider $H = H_0 + H_I$ where
\begin{eqnarray*}
  H_0 &=& H_b = \epsilon J_z \\
  H_I &=& H_a + H_{ab} = a^\dag a + \frac{2 \lambda}{\sqrt{N}}\left(a^\dag + a\right) J_x,
\end{eqnarray*}
with the Hamiltonian $H_a$ denoting subsystem $a$ (i.e. photon field), $H_b$ denoting subsystem
$b$ (i.e. matter) and $H_{ab}$ denoting their interaction.  The size-consistent \cite{Becker} effective Hamiltonian
in subsystem $a$ is
\begin{displaymath}
  H^{eff}_a = -\frac{1}{\beta}\langle e^{-\beta \left( H_I + L_0 \right)} - 1 \rangle^c_b.
\end{displaymath}
The index $c$ denotes cumulant averaging \cite{Becker} and the thermal average is carried out with respect to %%@
the matter degrees of freedom, i.e. $\langle A \rangle_b = {\rm Tr}_b \left(e^{-\beta \epsilon J_z}A \right)/{\rm %%@
Tr}_b\left(e^{-\beta \epsilon J_z} \right)$. The Liouvillian
superoperator $L_0$ is defined by $L_0 A = [H_b,A]$. The cumulants can be expanded in a series:
\begin{displaymath}
\langle e^{-\beta \left( H_I + L_0 \right)} - 1 \rangle^c_b =
\sum_{\nu = 1}^{\infty} \frac{(-\beta)^\nu}{\nu!} \langle (H_I + L_0)^\nu \rangle^c.
\end{displaymath}
We calculate the first two cumulants exactly:
\begin{eqnarray*}
  \langle H_I + L_0 \rangle^c &=& a^\dag a \\
  \langle (H_I + L_0)^2 \rangle^c &=& \lambda^2 (a+a^\dag)^2.
\end{eqnarray*}
When we calculate higher-order cumulants, and retain terms which are linear and quadratic in $a$ and $a^\dag$, we can derive a general
expression for every even and odd cumulant:
\begin{eqnarray*}
  \langle (H_I + L_0)^{2n} \rangle^c &=& \lambda^2 \epsilon^{2n - 2}(a+a^\dag)^2 \\
  \langle (H_I + L_0)^{2n + 1} \rangle^c &=& \lambda^2 \epsilon^{(2n+1)-2}(a+a^\dag)^2 \tanh\left(\frac{\beta %%@
\epsilon}{2}\right).
\end{eqnarray*}
We find that only second-order terms are required to describe the salient features of
the Dicke model at zero-temperature, hence we will exclude higher-order terms from the discussion.  The effective Hamiltonian is therefore
\begin{displaymath}
  H^{eff}_a = a^\dag a - \frac{\lambda^2}{\epsilon} \tanh \left(\frac{\beta
  \epsilon}{2}\right)\left(a+a^\dag\right)^2.
\end{displaymath}
In the low-temperature limit $\beta \rightarrow \infty$, this reduces to
\begin{equation}\label{Eq:eff_h}
  H^{eff}_{a (T=0)} = \omega \left(a^\dag a + \frac{1}{2} \right) + \gamma \left( a^{\dag 2} + a^2 \right) - \frac{1}{2}
\end{equation}
with $\omega = 1 -\frac{2 \lambda^2}{\epsilon}$ and $\gamma = -\frac{\lambda^2}{\epsilon}$.
Hence the optical properties of the Dicke model can be mapped onto a degenerate parametric process in which a classical field interacts with a non-linear medium. Equation (\ref{Eq:eff_h}) belongs to a class of squeezing Hamiltonians with SU(1,1) symmetry which have been shown to exhibit a ground-state phase transition \cite{Gerry1}.

%%%%%%%%%%%%%%%%%%%%%%%%%%%%%%%% FIGURE1 %%%%%%%%%%%%%%%%%%%%%%%%%%%%%%%%%%
\begin{figure}[!htbp]
\resizebox{7cm}{!}{\includegraphics*{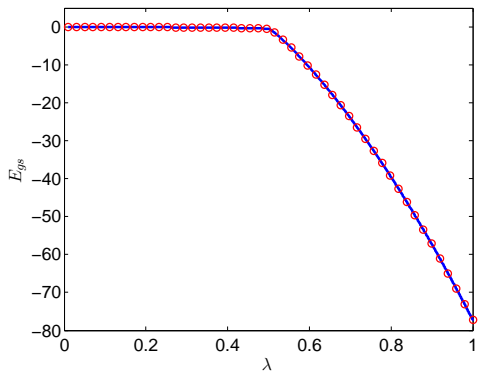}}
\caption{Ground state energy as a function of the light-matter coupling $\lambda$, as calculated with the size-consistent 
Hamiltonian of Eq. (\ref{Eq:eff_h}) (solid line) and the full Dicke model Hamiltonian (circles).}
\label{Fig:ge}
\end{figure}
%%%%%%%%%%%%%%%%%%%%%%%%%%%%%%%% FIGURE1 %%%%%%%%%%%%%%%%%%%%%%%%%%%%%%%%%%

We now show how our effective Hamiltonian not only reproduces the important features of the Dicke model, but also provides interesting optical signatures of its underlying quantum critical behavior: 
\vskip0.1in

\noindent {\it (i) Ground state energy, quantum critical point and scaling.} 
Figure \ref{Fig:ge} shows the excellent agreement between the exact numerical ground-state energy of the Dicke-model, and that obtained from Eq. (\ref{Eq:eff_h}).
In order to describe the underlying critical behavior, we perform a Bogoliubov 
transformation from the operators $a$,$a^\dag$ to squeezed $b$-bosons, such that 
$b^\dag = u a + v a^\dag$ and $|u|^2 - |v|^2 = 1$:
\begin{eqnarray*}
  b = \frac{a + \alpha a^\dag}{\sqrt{1-\alpha^2}} \, \, , 
  b^\dag = \frac{a^\dag + \alpha a}{\sqrt{1-\alpha^2}}
\end{eqnarray*}
and substitute for $a$,$a^\dag$ in Eq. (\ref{Eq:eff_h}).  The additive constant 
has no effect on the energy gap between
the ground and first excited states. Using 
$[b,b^\dag]=1$ yields
\begin{eqnarray}\label{Eq:bboson-h-eff}
  H^{eff}_{a (T=0)} &=& \frac{1}{1-\alpha^2}\bigg\{\left(\gamma \alpha^2 - \omega \alpha + \gamma \right)\left(b^{\dag %%@
2} + b^2\right) \nonumber \\
  & & \ \ \ + \left(\omega \alpha^2 - 4 \gamma \alpha + \omega \right)\left(b^\dag b + \frac{1}{2} \right)\bigg\}.
\end{eqnarray}
We choose $\alpha$ such that the coefficient of $(b^{\dag 2} + b^2)$ is
zero and apply the so-called resonance condition $\epsilon = 1$.  We find
\begin{displaymath}
  \alpha = -\frac{1-2 \lambda^2 \pm \sqrt{1 - 4 \lambda^2}}{2 \lambda^2}
\end{displaymath}
and the effective Hamiltonian (having arbitrarily chosen one of the two equivalent
solutions) becomes
\begin{equation}\label{Eq:bog_h_eff}
  H^{eff}_{a (T=0)} = -\sqrt{1-4 \lambda^2}\left(b^\dag b+\frac{1}{2}\right)\ .
\end{equation}
This effective Hamiltonian represents a simple harmonic oscillator in the
$b$-bosons and hence may be diagonalized by the number states of the
$b$-boson operators, i.e. $E_n = -\sqrt{1-4\lambda^2}\left(n+\frac{1}{2}\right )$.
The energy gap between the ground and first excited states is 
\begin{displaymath}
  \Delta E = \sqrt{1-4 \lambda^2}
\end{displaymath}
and hence we correctly deduce the quantum critical point as $\lambda = \lambda_c = 0.5$.  The energy gap is proportional to $\sqrt{\lambda_c - \lambda}$ as $\lambda
\rightarrow \lambda_c$, which is also in excellent agreement with previous results \cite{CMP}.  
The sub-radiant phase (i.e. $\lambda < \lambda_c$) 
is well described by the above Bogoliubov transformation -- the effective Hamiltonian
maps to a simple harmonic oscillator.  By contrast, in the super-radiant phase (i.e. 
$\lambda > \lambda_c$) the effective Hamiltonian in Eq. (\ref{Eq:bboson-h-eff}) resembles an inverted oscillator, as we discuss later.   
This finding highlights the inherent instability associated with the phase transition. 

\vskip0.1in

\begin{figure}[!htbp]
\resizebox{7cm}{!}{\includegraphics*{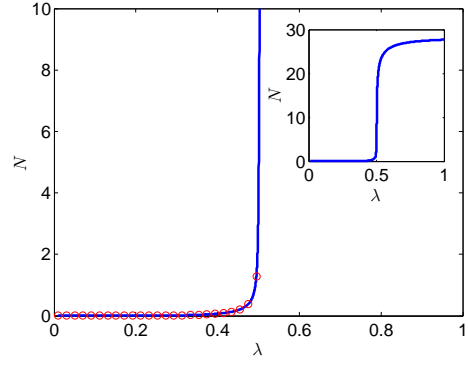}}
\caption{Analytical (circles) and numerical (solid line) calculation of the 
photon occupation number plotted as a function of $\lambda$. 
The inset shows the saturation of $N$ for large values of $\lambda$.}
\label{Fig:N}
\end{figure}

\begin{figure}[!htbp]
\resizebox{8cm}{!}{\includegraphics*{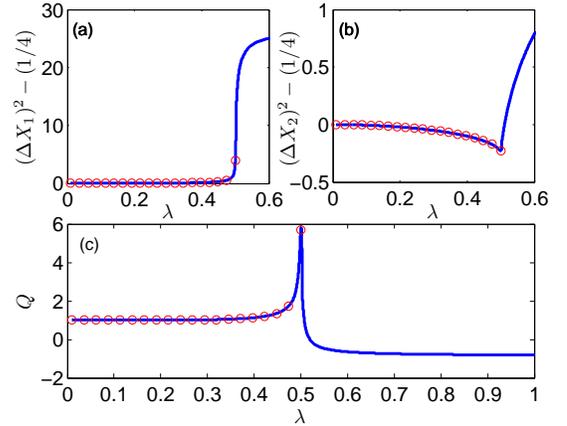}}
\caption{Analytical (circles) and numerical (solid line) 
calculations of (a) variance of $X_1$, (b) variance of $X_2$ and (c) 
Mandel $Q$-parameter vs. $\lambda$.}
\label{Fig:q-x1-x2}
\end{figure}

\noindent {\it (ii) Sub-radiant phase}. We obtain analytical expressions for the photon field occupation number $N$, the Mandel $Q$-parameter \cite{books} and the optical
squeezing \cite{books}, by exploiting the $SU(1,1)$ symmetry of the effective Hamiltonian (Eq. (\ref{Eq:eff_h})) \cite{Gerry1}. 
The Lie algebra of $SU(1,1)$ is generated by introducing the operators
\cite{Gerry1}
\begin{eqnarray*}
  K_0 &=& \frac{1}{2}\left(a^{\dag} a + \frac{1}{2} \right), \ \ 
  K_+ = \frac{1}{2}a^{\dag} a^{\dag}, \ \  K_- = \frac{1}{2}a a,
\end{eqnarray*}
with $[K_0,K_\pm]= \pm K_\pm, [K_{-},K_{+}] = 2 K_0$.
Equation (\ref{Eq:eff_h}) becomes
\begin{displaymath}
  H^{eff}_{a (T=0)} = 2 \omega K_0 + \frac{1}{2} \gamma (K_{+} + K_{-})
\end{displaymath}
with $\gamma$ redefined as  $\frac{-4 \lambda^2}{\epsilon}$.  
We introduce the $SU(1,1)$ coherent states 
\begin{displaymath}
  |\xi,k \rangle = \exp \left\{ z K_{+} - z^* K_{-}\right\}|0,k \rangle,
\end{displaymath}
where $z=-(\theta /2)e^{-i\phi}$ and $\xi = -\tanh(\theta/2)e^{-i \phi}$; $\theta$ and $\phi$ are group parameters with ranges $(-\infty,\infty)$ and
$[0,2\pi]$ respectively. The state $|0,k \rangle$  with $k = \frac{1}{4}$ is the 
vacuum squeezed state while $k=\frac{3}{4}$ corresponds to the squeezed
one-photon state. Now $N$, $Q$ and the squeezing can be calculated 
in terms of expectation values of the three generators.  For example
\begin{eqnarray*}
  \langle \xi,k|K_0|\xi,k \rangle &=& k \cosh \theta, \\
  \langle \xi,k|K_{\pm}|\xi,k \rangle &=& -k \sinh\theta e^{\pm i \phi}.
\end{eqnarray*}
The equations
of motion for $\theta$ and $\phi$ are
\begin{eqnarray*}
  \stackrel{.}{\theta} &=& -(k \sinh\theta)^{-1} \frac{\partial \mathscr{H}}{\partial \phi}, \\
  \dot{\phi} &=& (k \sinh\theta)^{-1} \frac{\partial \mathscr{H}}{\partial \theta},
\end{eqnarray*}
where $\mathscr{H} = \langle \xi,k| H^{eff}_{a (T=0)} |\xi,k \rangle$.  We hence obtain
\begin{eqnarray*}
  \mathscr{H} &=& 2 \omega k \cosh \theta - \gamma k \sinh \theta \cos \phi - \frac{1}{2}, \\
  \dot{\theta} &=& -\gamma \sin \phi, \\
  \dot{\phi} &=& 2 \omega - \gamma \coth \theta \cos \phi\ .
\end{eqnarray*}
The stationary points must satisfy $\dot{\theta} = \dot{\phi} = 0$, yielding 
$\phi = n \pi$ with $n=0,1,2,\dots$. If $n$ is odd, we find
\begin{displaymath}
  \theta = {\rm arccoth}\left(\frac{1-2\lambda^2}{2\lambda^2}\right)\ .
\end{displaymath}
Hence no real solution exists above the critical point $\lambda_c =
0.5$. As hinted at earlier with the suggestion of an inverted oscillator, this is because the energy $\mathscr{H}$ becomes unbounded above $\lambda_c$.  In the sub-radiant phase (i.e. $\lambda<\lambda_c$) the photon number $N=\langle \xi,k | a^{\dag}a| \xi,k \rangle $ 
can be written as
\begin{eqnarray*}
  N&=& 2\langle K_0 \rangle - \frac{1}{2}= 2 k \cosh \theta - \frac{1}{2}\ \ .
\end{eqnarray*}
Figure \ref{Fig:N} shows that this expression accurately reproduces the numerical results, implying that the field is effectively in a squeezed state in the sub-radiant phase. 
%%%%%%%%%%%%%%%%%%%%%%%%%%%%%%%% FIGURE2 %%%%%%%%%%%%%%%%%%%%%%%%%%%%%%%%%%
%%%%%%%%%%%%%%%%%%%%%%%%%%%%%%%% FIGURE2 %%%%%%%%%%%%%%%%%%%%%%%%%%%%%%%%%%
\noindent We now turn to the optical squeezing itself, which is defined in terms of the quadrature operators $X_1 = \frac{1}{2}(a+a^{\dag})$ and 
$X_2 = \frac{1}{2i}(a-a^{\dag})$ with $[X_1,X_2]=i/2$. This yields $(\Delta X_1)^2(\Delta X_2)^2\geq %%@
\frac{1}{16}$ where 
$(\Delta X_i)^2=\langle X_i^2\rangle-\langle X_i\rangle^2$.  Hence squeezing exists if $(\Delta X_i)^2\leq \frac{1}{4}$.
We obtain
\begin{eqnarray*}
  (\Delta X_{1,2})^2 &=& k\left(\cosh \theta \mp \cos \phi \sinh \theta\right).
\end{eqnarray*}
For $k=\frac{1}{4}$ and $\phi=n\pi$ with $n=0,1,2,\dots$, we find that $(\Delta X_{2})^2\leq \frac{1}{4}$ throughout the sub-radiant phase 
(see Fig. \ref{Fig:q-x1-x2}) and the ground state is a minimum-uncertainty squeezed state, i.e. $(\Delta X_1)^2(\Delta X_2)^2 = \frac{1}{16}$.
The Mandel $Q$-parameter in the sub-radiant phase is given by
\cite{books}
\begin{eqnarray*}
  Q &=& \frac{\langle a^{\dag} a a^{\dag} a \rangle - \langle a^{\dag} a \rangle^2}{\langle
  a^{\dag} a \rangle} - 1 
    = \frac{4\langle K_{0}^{2} \rangle - 4\langle K_0 \rangle^2}{2\langle
    K_0 \rangle - \frac{1}{2}} - 1 \\
    &=& \frac{k\left((1+2k)\cosh(2\theta)+2k-1 \right)-4k^2 \cosh^2 \theta}{2k\cosh \theta - \frac{1}{2}}-1.
\end{eqnarray*}
which, for $k=\frac{1}{4}$, is plotted in Fig. 
\ref{Fig:q-x1-x2}(c).

%%%%%%%%%%%%%%%%%%%%%%%%%%%%%%%% FIGURE3 %%%%%%%%%%%%%%%%%%%%%%%%%%%%%%%%%%
%%%%%%%%%%%%%%%%%%%%%%%%%%%%%%%% FIGURE3 %%%%%%%%%%%%%%%%%%%%%%%%%%%%%%%%%%
\vskip0.1in

\noindent {\it (iii) Super-radiant phase.} We
introduce the canonical position and momentum operators
\begin{equation}
  \hat{x} = \frac{1}{\sqrt{2 \omega}} (a + a^\dag);\  \hat{p} = \frac{1}{i}\sqrt{\frac{\omega}{2}}(a-a^\dag)
\end{equation}
where $[\hat{x},\hat{p}] = i$.  Equation (\ref{Eq:eff_h}) becomes
\begin{equation}\label{Eq:eff_ho_h}
  H^{eff}_{a (T=0)} = \frac{\hat{p}^2}{2 \bar{m}}+\frac{1}{2}\bar{m}\bar{\omega}^2 \hat{x}^2 ,
\end{equation}
with
\begin{displaymath}
  \bar{m} = \frac{1}{1-\frac{2 \gamma}{\omega}} \ \ \ \ \ ,  \qquad \bar{\omega}^2 =  \omega^2\left(1+\frac{2
  \gamma}{\omega}\right)\left(1-\frac{2 \gamma}{\omega}\right)\ \ \  ,
\end{displaymath}
where we have again ignored the additive $\frac{1}{2}$.  
%%%%%%%%%%%%%%%%%%%%%%%%%%%%%%%% FIGURE4 %%%%%%%%%%%%%%%%%%%%%%%%%%%%%%%%%%
\begin{figure}[!htbp]
\resizebox{7.5cm}{!}{\includegraphics*{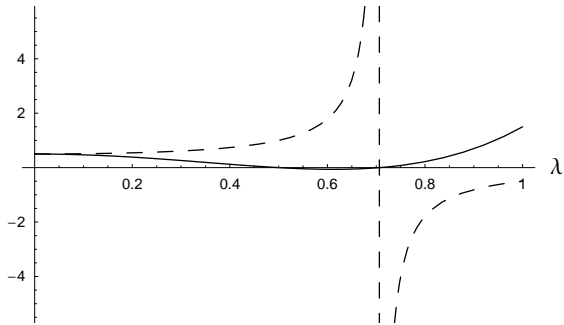}}
\caption{Coefficients of $\hat{x}^2$  (solid line) and $\hat{p}^2$ (dashed line) in Equation (\ref{Eq:eff_ho_h}) vs.
$\lambda$.}
\label{Fig:x-p}
\end{figure}
%\includegraphics[width=0.45\textwidth]{figure4.eps}
%%%%%%%%%%%%%%%%%%%%%%%%%%%%%%%% FIGURE4  %%%%%%%%%%%%%%%%%%%%%%%%%%%%%%%%%%
The coefficients of 
$\hat{x}^2$ and $\hat{p}^2$ are plotted against $\lambda$ in Fig. \ref{Fig:x-p}. These plots show that in the sub-radiant phase (i.e. $\lambda < 0.5$) the system
is equivalent to a harmonic oscillator and as such is square-integrable -- hence it can be
diagonalized using basis states of the number operator $a^\dag a$.
In the super-radiant phase (i.e. $\lambda > 0.5$) the system becomes an
{\em inverted} harmonic oscillator -- first in momentum, and then in position.  The properties
of an inverted potential harmonic oscillator are discussed elsewhere
\cite{Barton}.  It is interesting to note that the energy becomes unbounded
and the energy eigenspectrum becomes continuous -- in addition, the inverted
potential harmonic oscillator has been used as a model of instability in
relation to quantum chaos \cite{ho-chaos}.  This confirms the claim of Ref. \cite{brandes-chaos} that the system undergoes a transition from quasi-integrable to quantum chaotic behavior at the quantum critical point.

In the
super-radiant phase, the system cannot be described in a simple analytic form -- except in the limit $\lambda \rightarrow \infty$. However the
quantities in which we are interested can still be calculated using numerical simulations.
These numerical simulations suggest that -- with the exception of the energy -- each
quantity of interest rapidly converges to a single value as $\lambda$ is
increased.  Remarkably, we find that only a small number of photons are required to obtain extremely accurate results.  Furthermore, the scaling remains intact as we increase the system size.
Figure \ref{Fig:N} shows that the occupation number $N$ is initially zero, and remains small but finite as we increase $\lambda$ toward $\lambda_c$.  At this critical point, the occupation becomes 
macroscopic, i.e. the system ceases to be sub-radiant and enters the super-radiant phase.  
The behaviour seen in Fig.\ref{Fig:N} compares extremely well with the 
properties predicted for the output photon flux in Ref. \cite{Carmichael}.  Indeed Fig. \ref{Fig:N} (inset) suggests that as $\lambda$  increases further,
the occupation converges toward a single value. 
Figures \ref{Fig:q-x1-x2}(a) and (b) indicate that the radiation field survives in an {\it ideal} 
squeezed state up to the critical point $\lambda_c=0.5$, but that this squeezed state then breaks down in the super-radiant phase. At  $\lambda_c$, a discontinuity appears in $(\Delta X_{2})^2$ which is associated with a sudden change in the photon statistics -- a finding confirmed by the Mandel $Q$-parameter (Fig.\ref{Fig:q-x1-x2}(c)). In the
sub-radiant phase (i.e. $\lambda<\lambda_c$) we have $Q>0$ indicating that the photon statistics are super-Poissonian.  At $\lambda_c $, $Q$ diverges.
Above $\lambda_c$, $Q<0$ which indicates that the statistics are
sub-Poissonian.  As $\lambda$ is increased further, $Q \rightarrow -1$
and hence $Q$ takes on its most negative possible value, i.e. the multi-photon state becomes a Fock state \cite{books}.

To summarize, we have analyzed a known Quantum Phase Transition from the entirely new perspective of the accompanying photon field. The predicted optical signatures should be readily measurable in either atom-cavity or nanostructure-cavity systems \cite{CMP,Brandes,Carmichael} using existing optical techniques \cite{books}. 

We are extremely grateful to Luis Quiroga and Jos\'e Reslen for earlier discussions about this work.
A.O-C. thanks Trinity College, Oxford for financial support.

\end{document}